\documentclass[a4paper]{spie}

\usepackage{amsmath,amsfonts,amssymb}
\usepackage{graphicx}
\usepackage[colorlinks=true, allcolors=blue]{hyperref}
\usepackage{times}
\usepackage{mathptmx}
\usepackage{xspace}

\title{The BlueMUSE data reduction pipeline:
       lessons learned from MUSE and first design choices}
\author[a]{Peter M.\ Weilbacher}
\author[b]{Sven Martens}
\author[c]{Martin Wendt}
\author[a]{Martin M.\ Roth}
\author[b]{Stefan Dreizler}
\author[a]{Andreas Kelz}
\author[d]{Roland Bacon}
\author[d]{Johan Richard}
\affil[a]{Leibniz-Institut für Astrophysik Potsdam (AIP),
          An der Sternwarte 16, 14482 Potsdam, Germany}
\affil[b]{Institut für Astrophysik, Friedrich-Hund-Platz 1,
          37077 Göttingen, Germany}
\affil[c]{Institut für Physik und Astronomie, Universität Potsdam,
          Karl Liebknecht-Str.\ 24/25, 14476 Golm, Germany}
\affil[d]{Univ Lyon, Univ Lyon1, Ens de Lyon, CNRS, Centre de Recherche
          Astrophysique de Lyon UMR5574, F-69230, Saint-Genis-Laval,
          France}
\authorinfo{Send correspondence to PMW via e-mail to
            \href{mailto:pweilbacher@aip.de}{pweilbacher@aip.de}.}

\def\arcmin{\hbox{$^\prime$}}

\def\farcs{\hbox{$.\!\!^{\prime\prime}$}}

\begin{document}
\maketitle

\begin{abstract}
BlueMUSE is an integral field spectrograph in an early development stage for
the ESO VLT.
For our design of the data reduction software for this instrument, we are first
reviewing capabilities and issues of the pipeline of the existing MUSE
instrument. MUSE has been in operation at the VLT since 2014 and led to
discoveries published in more than 600 refereed scientific papers.
While BlueMUSE and MUSE have many common properties we briefly point out a
few key differences between both instruments.
We outline a first version of the flowchart for the science reduction, and
discuss the necessary changes due to the blue wavelength range covered by
BlueMUSE. We also detail specific new features, for example, how the pipeline
and subsequent analysis will benefit from improved handling of the data
covariance, and a more integrated approach to the line-spread function, as
well as improvements regarding the wavelength calibration which is of extra
importance in the blue optical range.
We finally discuss how simulations of BlueMUSE datacubes are being implemented
and how they will be used to prepare the science of the instrument.
\end{abstract}

\section{INTRODUCTION}\label{sec:intro}
BlueMUSE is a new project that is building on the legacy of the successful MUSE
instrument\cite{Bacon+10,2014Msngr.157...13B} at the ESO VLT. It has just
started the pre-design (``Phase A'') and is planned to be installed at one of
the VLTs in 2030. The science cases to build this new instrument and its
planned top-level parameters were already outlined in a White
Paper\cite{BlueMUSE_1906.01657}.

Briefly, BlueMUSE will be a panoramic integral field spectrograph which is
thought to cover at least $1\arcmin\times1\arcmin$ on the sky with fine
sampling of 0\farcs3/pix or better.  It will cover the blue part of the optical
spectrum, from the atmospheric cutoff at 350\,nm to about 550\,nm in one shot
with high throughput. It will feature twice the spectral resolution of MUSE,
i.e.\ $\left<R\right>\approx3600$ when averaged over the wavelength range. Like
MUSE, BlueMUSE will be made of 24 slicer-style integral field units and record
the spectra on as many 4k$\times$4k CCDs. BlueMUSE will not, however, be
coupled to an adaptive optics system, it will therefore operate with a single
mode, with a stronger focus on instrument stability and even higher operational
efficiency with minimal overheads.

Since both instruments have a very similar design, it makes sense to look at
the successes and failures of the MUSE pipeline\cite{musepipeline}
(Sect.~\ref{sec:museissue}) before starting to design a pipeline for the new
BlueMUSE instrument (in Sect.~\ref{sec:bluemuse}). We will discuss selected
items of the new design in a bit more detail (Sect.~\ref{sec:bmse_new}) and
present the basic ideas of the code to create simulated datacubes
(Sect.~\ref{sec:sim}).

\section{MUSE PIPELINE: LESSONS LEARNED}\label{sec:museissue}
The MUSE instrument and its pipeline was a great success.
As of June 2022 about 625 refereed publications are at least partially based on
its data. In 2021 MUSE enabled the most publications of any ESO instrument,
surpassing even HARPS and UVES.\footnote{See
\url{https://www.eso.org/sci/php/libraries/pubstats/\#vlt} for ESO publication
statistics.} In part, this success was due to a data reduction pipeline that
could deliver science-ready datacubes with standard parameters for a large
range of science topics, ranging from solar-system
objects\cite{2019Icar..331...69I,2019A&A...628A.128O}, Galactic
nebulae\cite{McLeod15b,2020A&A...634A..47M} and
clusters\cite{2018MNRAS.473.5591K,2019A&A...632A...3G},
nearby\cite{2018A&A...618A...3R,2020A&A...633A..79K,2022A&A...659A.191E} as
well as high-redshift\cite{2018Natur.562..229W,2021A&A...647A.107B} galaxies.

In fact, that the pipeline was ready to support already the first commissioning
run in February 2014 was of great importance. It could at that time already run
the full suite of reduction steps to produced a datacube with most of the
instrument signature and atmospheric effects removed. This allowed the pipeline
team to concentrate on the most visible deficiencies, like an inverted field of
view, a complete rewrite of the skyflat handling and of the computation and
application of the line-spread function.
All this resulted in the first MUSE science papers\cite{2014MNRAS.445L..79E,2014MNRAS.445.4335F}
getting written and submitted within a month of the science verification
observations that they were based on.

\subsection{Pipeline design}
\begin{figure}[ht]
\begin{center}
\begin{tabular}{c}
\includegraphics[height=4.9cm]{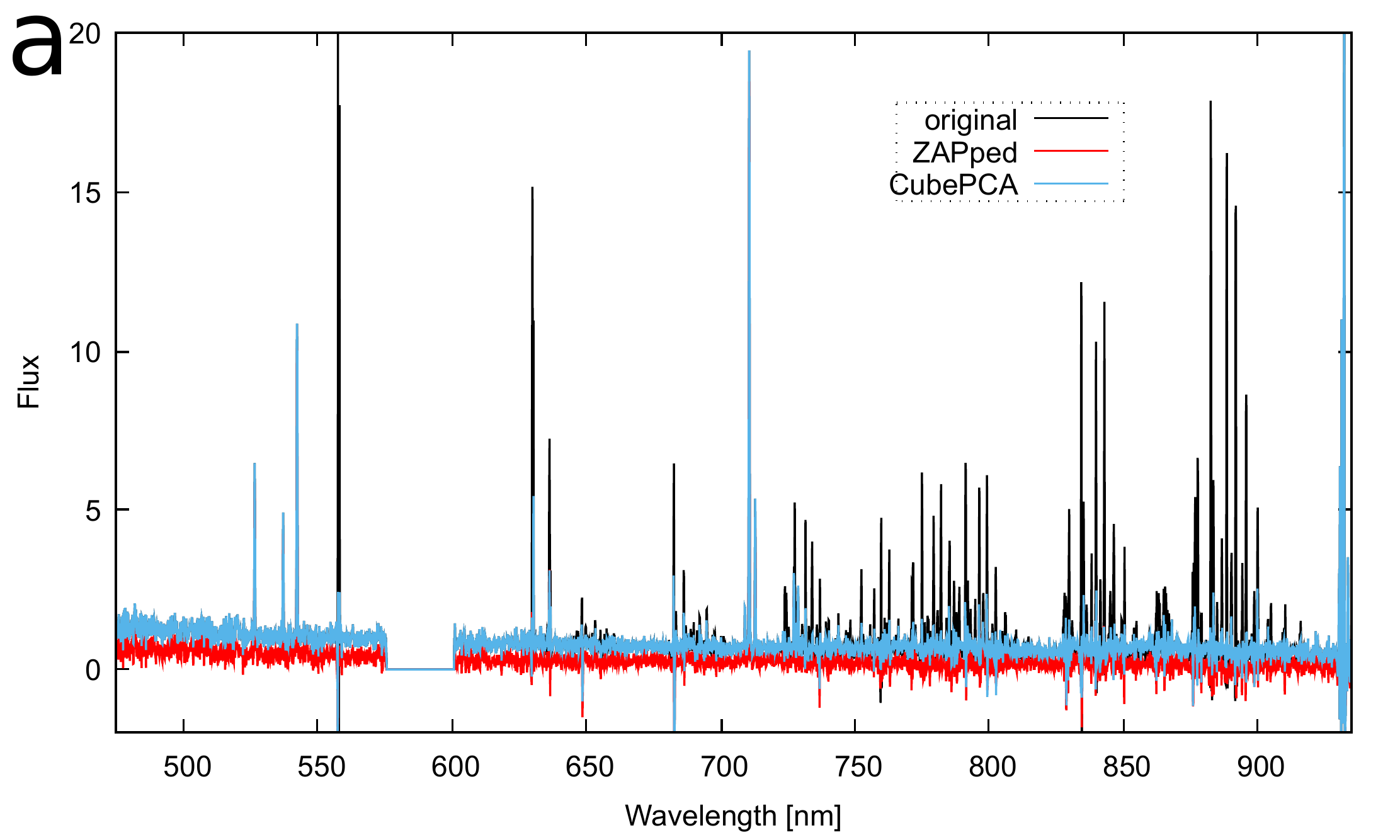}
\includegraphics[height=4.9cm]{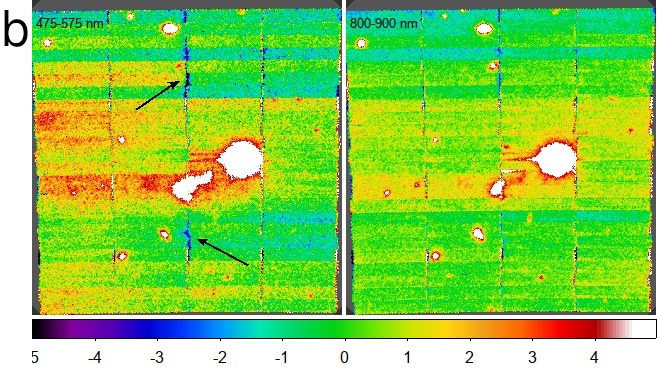}
\end{tabular}
\end{center}
\caption{MUSE data issues.
  \textbf{(a)}:
  HII region spectrum with pipeline sky subtraction (black), with
  additional ZAP cleaning (red), and processed with CubePCA (blue). While ZAP
  cleans the sky residuals, it also changes the continuum level. CubePCA leaves
  the continuum intact but also misses some of the telluric artifacts.
  \textbf{(b)}:
  The picture shows images created by integrating one MUSE cube (containing a
  field with one bright star and a few small galaxies) over 100\,nm at
  the blue and red end of the wavelength range. Each of the 24 IFUs is visible
  by a 12\,pix high band across the field of view, the relative levels vary with
  wavelength. Arrows mark two strong dark features at the intersection of two
  slicer stacks.
}\label{fig:skysub_ff}
\end{figure}

However, a number of issues are still present in the pipeline and/or in the
reduced data. For example, while the pipeline is parallelized in a large
extent, only half of its processing is done with OpenMP parallelization while
the other half essentially relies on the user starting multiple processes in
parallel. This was an early design choice, imposed by the processing
environment that was thought to be realized in about 2008 but was not present
(any more) when the instrument was commissioned in 2014. This division resulted
in a somewhat artificial split into ``basic''- and ``post''-processing which
also made it necessary to save large files to disk in between. Effectively,
this approach costs extra (temporary) diskspace\footnote{about 10 GiB per MUSE
  exposure, compared to raw data size of 0.8 GiB and final datacube of 2.7 GiB}
and adds processing time (only $\sim\,3$\%),
but also complexity because the correct input files for the next step need to
be determined (sometimes by user interaction).

\subsection{Issues: sky subtraction}
The most obvious issue with the data produced by the pipeline was related to
the sky subtraction: the bright sky lines left unexpectedly large residuals.
Various groups of MUSE users therefore developed separate packages to handle
telluric emission lines better, ZAP\cite{2016MNRAS.458.3210S},
CubEx\cite{2019MNRAS.483.5188C}, and CubePCA\cite{2022A&A...659A.124H}, among
the most prominent. This was especially necessary for science programs that took
``deep'' exposures of mostly empty sky regions which then tried to find faint
line-emitting galaxies using automated methods\cite{2015A&A...575A..75B,2019A&A...624A.141U}.
An additional complication for these extra procedures was that the
exposure-combination algorithm implemented in the MUSE pipeline could no longer
be used, since the extra sky-subtraction procedure worked only on resampled 3D
cubes and not on the (unresampled) intermediate data. On the other hand, these
extra processing steps could not be implemented in the pipeline itself, because
they only work well, if most of the exposure is empty, which is only the case
for a small fraction of the MUSE exposures, and because PCA-methods are hard to
adapt to data that are not sampled on a regular grid, like intermediate MUSE
data.
Finally, while these extra sky-subtraction programs very efficiently suppress
telluric residuals, they often change the data in unforeseen ways. An example
can be seen in Fig.~\ref{fig:skysub_ff}a where both extra processing steps
remove the sky artifacts, but ZAP also removes the continuum of the HII region
while CubePCA does not.

\subsection{Issues: flat-fielding}
The second obvious issue are patterns visible in images created from the
reduced datacubes by integrating over many wavelength planes
(Fig.~\ref{fig:skysub_ff}b).  MUSE is built from 24 individual slicer IFUs which
cover the full width of the field.\footnote{The layout is described in detail
  in the MUSE User Manual \cite{MUSE_User_Manual_v10.4} and the pipeline
  paper\cite{musepipeline}.}
They visibly stand out in all exposures, and their pattern of dark and bright
change with wavelength. On a smaller level, the four stacks of slices seem to
create different patterns, especially between the two slicer stacks on the left
and on the right of each IFU. Within each slicer stack, the images show an
almost constant level. Finally, on the smallest scales, the horizontal
intersections between the slicer stacks show a few especially dark (black
arrows in Fig.~\ref{fig:skysub_ff}b) and/or bright pixels.

While both issues, strong sky subtraction residuals and integrated image
artifacts, are very different phenomena and their root cause is not finally
understood, they are both strongly tied to the quality of flat-fielding.
Indeed, several calibrations (line-spread function, wavelength calibration,
OH-transition flux ratios, and line flux homogeneity) were tested as to their
effects on the sky subtraction accuracy.
It turned out that the fluxes are varying across the field at the $\sim\!$1.7\%
level, which is about the same level of variations seen in the sky background
(as in Fig.~\ref{fig:skysub_ff}b). The sky-line residuals visible in
Fig.~\ref{fig:skysub_ff}a are almost all less than 1\% of the original flux of
the line (this was shown to be true for most spectra\cite{musepipeline}). The
assumed flux ratios and the wavelengths of the telluric emission lines have
systematic errors, too, but their effects are typically below this level.

Therefore, by improving the flux calibration across the field through better
flat-fielding, one could also improve the accuracy of the sky subtractions.
For BlueMUSE, however, the only bright telluric emission lines in the
wavelength range will be [O\textsc{I}]5577 and the [N\textsc{I}]5200 doublet.
Sky subtraction will therefore be much less critical. But even to subtract
fainter emission lines that would otherwise hamper detection of faint line
emitters, we will strive to provide good correction for telluric emission.

\subsection{Remedies}
Under the assumption that the sky background in a MUSE exposure is really
constant across the small $1'\times1'$ field, two important approaches were
used to flatten the data further. The first computed the integrated sky
brightness in several wavelength ranges for each slice in the MUSE field, and
compared this to the average sky brightness. Based on this, correction factors
were computed. After applying them, only small gradiants along the slices and
the "gaps" between the slicer stacks were visible in images created from the
cubes. This approach was used for many MUSE datasets where most of the exposure
was "empty", i.e., in extragalactic fields\cite{2017A&A...608A...1B,2019A&A...624A.141U}.
For fields of nearby large objects, this approach is not feasible.

The "gaps" between the slicer stacks were often masked out before combining
exposures\cite{2017A&A...608A...1B,2019A&A...624A.141U}. Because this resulted
in reduced S/N in these regions, the idea of a "super-flat" was more recently
used to improve S/N and cosmetic quality of extremely deep
exposures\cite{2021A&A...647A.107B}.  For this super-flat, a large number of
almost empty exposures have to be (median-)combined, onto the same grid as the
exposure to be improved, but without taking into account their original dither
offsets. This creates a datacube that has almost exactly the same artifacts as
the science exposures, applying it to the science exposure then almost
completely removes them, without affecting the scientific content of the
exposure. It could, in principle, also be applied to exposures of large
objects, if enough exposures without large offsets were taken with MUSE in the
days around it. The drawback of this approach is the huge computational effort
to create such a superflat for each science exposure.

For BlueMUSE, we are studying how to improve the flat-fielding so that these
issues do not arise. Improved instrument stability of BlueMUSE will help in
this regard.

\section{BlueMUSE PIPELINE: FIRST IDEAS}\label{sec:bluemuse}

\begin{figure}[ht]
\begin{center}
\begin{tabular}{c}
\includegraphics[width=0.5\linewidth]{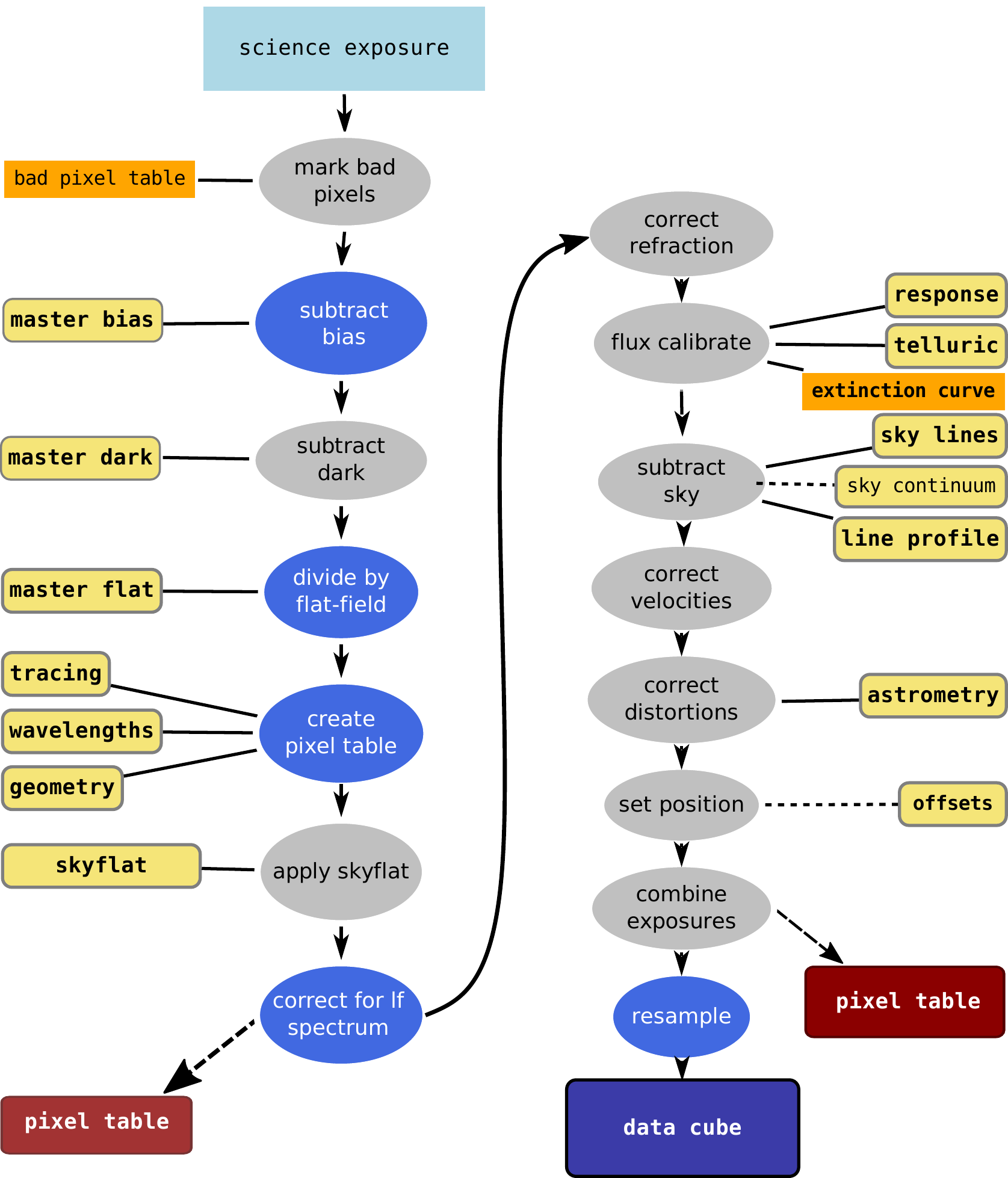}
\end{tabular}
\end{center}
\caption{An idea for the scientific reduction algorithm for BlueMUSE.
}\label{fig:flowchart}
\end{figure}

As all ESO pipelines, the BlueMUSE pipeline will be embedded into the ESO data
flow system, i.e., it will be implemented as part of CPL/HDRL plugins to the
EsoRex\cite{2015ascl.soft04003E} and EDPS\cite{2022SPIE_EDPS} systems.

Based on the MUSE pipeline, we show a first flowchart for the science reduction
of the BlueMUSE pipeline in Fig.~\ref{fig:flowchart}. Since the instrument will
be very similar, most of the steps will remain the same, from bad pixel masking
using an external file, over bias subtraction, optional dark subtraction, and
CCD-based flat-fielding using lamp-flat exposures. We then convert the images
into tables of pixels using information on the location of the spectra (the
tracing info), the wavelength calibration, and the instrument geometry. Onto
these we apply the secondary sky-flat correction and correct the data for the
overall shape of the lamp-flat spectrum, before correcting the data for
atmospheric refraction. The refraction correction will likely work well using
standard formulae and environmental measurements from the telescope telemetry
in the FITS header.

The next step will apply the flux calibration using a response curve derived
from a spectrophotometric standard, and apply an atmospheric extinction
correction.  This extinction correction could be derived from standard stars as
well, if several were taken during photometric conditions in the same night at
different airmasses.
The flux calibration will include the possibility to correct telluric
correction, even though no significant molecular absorption bands are expected
to be visible in the wavelength range of BlueMUSE.\footnote{The ozone bands in
  the UV-blue range are usually too weak and too broad and hence often not
  corrected in astronomical spectroscopy. Any spectrum to be used as correction
  could be input from external tools like ESO
  \texttt{molecfit}\cite{molecfit,2015ascl.soft01013S}.}
The sky subtraction, distortion correction and astrometric position correction,
as well as the exposure combination will work in the same way for BlueMUSE as
they did for MUSE. The correction to (barycentric) velocity can be improved
from recent work for other ESO pipelines (e.g., ESPRESSO), based on code from
the SOFA\cite{1996ASPC..101..207W,2014ascl.soft03026I} / ERFA\footnote{ERFA is
  a re-licensed version of SOFA and available at
  \url{https://github.com/liberfa/erfa}.}
initiative.

While saving of the pixel tables is no longer necessary at any step, we foresee
the \emph{option} of saving the intermediate files after all, to let users
optimize the data in different ways, even using code outside the pipeline
proper. All steps of the science reduction of BlueMUSE will be executable as
one integrated pipeline and as finegrained steps using the same library
functions for testing and debugging.

As for MUSE, we will delay the sampling of the data into the regular grid of a
datacube only as the very last step of the science reduction. But for BlueMUSE
we will take into account the covariance (see Sect.~\ref{subsec:covar}) and also
propagate the spectral line profile throughout of resampling.

\section{BlueMUSE PIPELINE: SPECIFIC IMPROVEMENTS}\label{sec:bmse_new}
We have so far studied at a few specific items that we will implement in a
different way in the BlueMUSE pipeline. We will discuss three of them in this
section.

\subsection{Wavelength calibration}\label{subsec:wavecal}
One of the main differences with regard to the MUSE instrument is that the
wavelength calibration using arc lamps will not work as well in the blue
wavelength range. The calibration unit (CU)\cite{2022SPIE_BlueMUSE_CU} will
therefore be using a Fabry-Perot system or a laser frequency comb as the light
source to densely populate the spectral range. This will create a near-equal
spaced lines on the detector, so that the association of lines to reference
wavelengths will not work with the pattern matching implemented
before\cite{musepipeline}.  Instead, we will create an initial zero-order
calibration based on a standard arc-lamp exposure, and then construct the fully
sampled wavelength solution using all lines from the comb or Fabry-Perot
system. To save space, the full wavelength solution will be saved as a
two-dimensional polynomial for each slice of the BlueMUSE instrument.

\subsection{Line-spread function}\label{subsec:lsf}
\begin{figure}[ht]
\begin{center}
\begin{tabular}{c}
\includegraphics[width=0.8\linewidth]{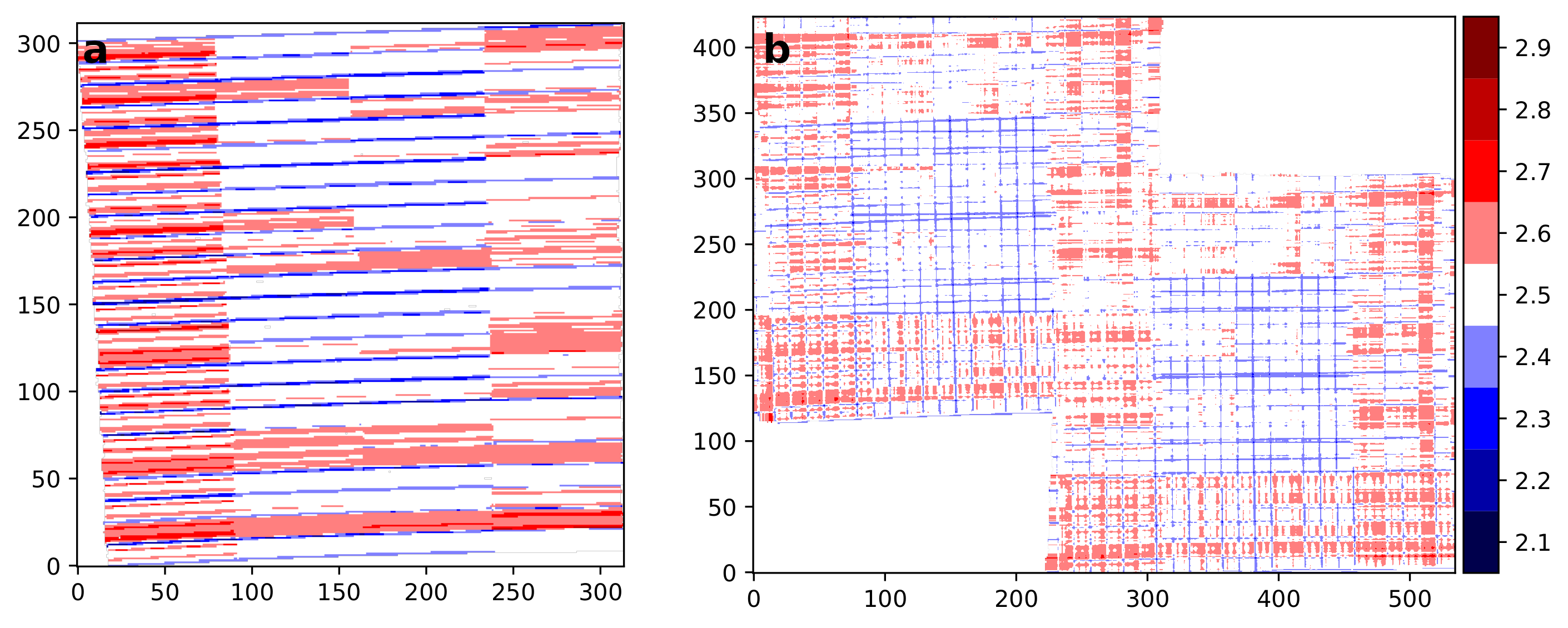}
\end{tabular}
\end{center}
\caption{Spatially resolved line-spread function FWHM over the MUSE field.
  \textbf{(a)}:
  LSF at 657.4\,nm propagated to the cube of a single MUSE exposure.
  The FWHM varies between about 2.1 and 2.9\,\AA\ width.
  \textbf{(b)}:
  LSF at the same wavelength, but for a MUSE cube created from six partially
  overlapping exposures, taken at 0, 90, and 180$^\circ$ angles. The variations
  were smoothed out and range only between about 2.3 and 2.7\,\AA\ FWHM.
}\label{fig:lsf}
\end{figure}

Using the Fabry-Perot setup of the BlueMUSE CU, we should be able to improve on
the estimation of the line-spread function (LSF) that was based only on a few
bright arc lines in the MUSE instrument. We are planning to use scanning
techniques to sample the LSF at better resolution than single arc lines allow.
We will therefore be able to derive the LSF not only for each slice of the
instrument, but for even for spatial positions along the slices.

The LSF will be used within the pipeline to improve the sky subtraction. But to
make it available to users of the instrument for the science analysis we will
also propagate it through the data reduction and create a datacube of the
instrumental profile (at least in terms of its FWHM). First tests doing that
were conducted with the MUSE pipeline for existing data, the results are shown
in Fig.~\ref{fig:lsf}. We can see that the pattern in a single exposure
(Fig.~\ref{fig:lsf}a) clearly shows the pattern of the slices, which are the
smallest elements for which we know the LSF. When combining more exposures that
were taken with rotational and positional offsets, the differences are smoothed
out (see Fig.~\ref{fig:lsf}b). In the analysis, LSF information like this can
be further propagated through, e.g., Voronoi binning\cite{2003MNRAS.342..345C}
to be used for stellar population fits with tools like
\texttt{pPXF}\cite{2017MNRAS.466..798C}.

\subsection{Covariance propagation}\label{subsec:covar}
\begin{figure}[ht]
\begin{center}
\begin{tabular}{c}
\includegraphics[width=0.75\linewidth]{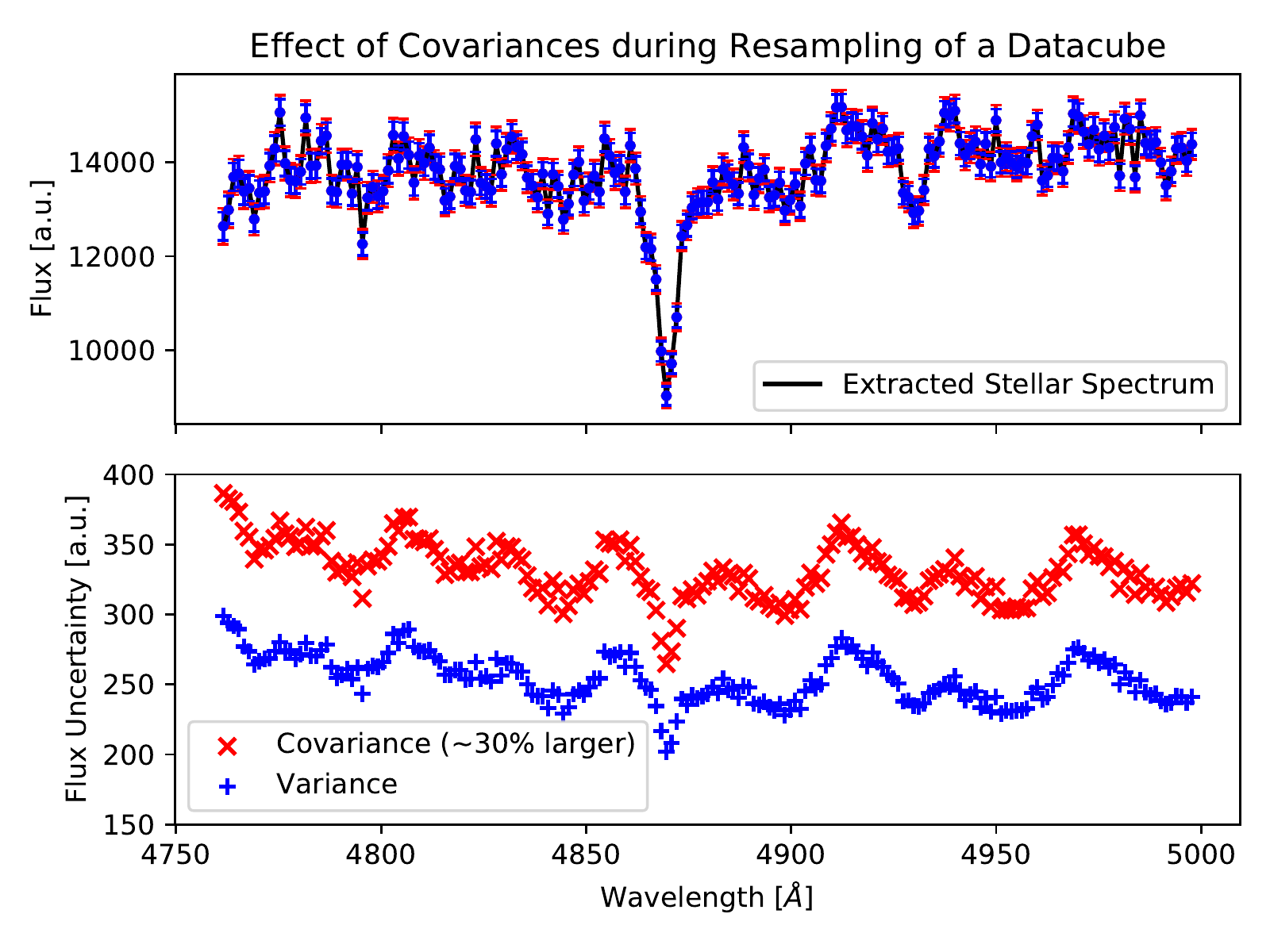}
\end{tabular}
\end{center}
\caption{Stellar spectrum extracted from a partial MUSE observation with the
         estimated flux uncertainties, which were calculated with
         (\textcolor{red}{red}) and without (\textcolor{blue}{blue}) taking the
         covariances into account. It is apparent that in this case the
         uncertainties are approximately 30\% larger when taking covariances
         into account.
}\label{fig:covar}
\end{figure}

One part of the MUSE pipline is resampling the data from an irregular to a
regular grid to create a datacube object\cite{musepipeline}. This process
causes variance values of single pixels to be transferred into covariances. The
pipeline authors argue\cite{musepipeline} that it would be infeasible to carry
these covariances through further pipeline steps due to the size of the data
involved. Since the MUSE pipeline only resamples the data once, the variances
per voxel of the output datacube are still accurate. However, combining
multiple voxels from a datacube in further analyses beyond the scope of the
reduction pipeline without considering the involved covariances could cause
significant underestimation of the uncertainties of flux values.

While it is true that the full covariance matrix would be too large to handle,
it turns out that most entries of this matrix are zero. Each resampling method
implemented in the MUSE pipeline only considers pixel values that are in some
spatial and spectral proximity to the position of the output voxel. Therefore,
covariances of a voxel are only offset from zero for other voxels in its
proximity. With this in mind, the number of entries that are non-zero is likely
small enough to be manageable for a whole datacube.

To put this to the test we implemented the drizzle-like resampling from the
MUSE pipeline. We then added the handling of covariances according to the
approach used for the pipeline of the Mapping Nearby Galaxies at Apache Point
Observatory (MaNGA)\cite{2016AJ....152...83L}. We applied this algorithm to a
spatially and spectrally limited part of a MUSE observation and extracted the
spectrum of a single star from the reduced datacube. This spectrum is shown in
the upper panel of Fig.~\ref{fig:covar}. The flux uncertainties that are shown
in that panel and the lower panel were calculated with and without taking the
covariances into account. It is apparent that uncertainties derived only from
variances are about 30\% smaller than the ones derived by including the
covariances.

The next step is to implement this method in the existing MUSE pipeline and run
it on a full MUSE observation. While we expect that the additional computation
time cost is not significant, this still needs to be tested. Based on the
reduction with covariances described above, we anticipate that the additional
storage size needed is in the order of $10\,$GB for a full MUSE observation
using the drizzle-like resampling method. This estimate would change for other
resampling methods that take into account pixels that are further away from the
output voxel. Overall, this seems to be a promising approach to account for
covariances during reduction with the MUSE pipeline.

\section{BlueMUSE SIMULATIONS}\label{sec:sim}
In the late preparation phase before first light of MUSE, the consortium worked
on a 'QSIM' tool to produce synthetic data cubes reflecting targeted science
observations.
These were intended to allow the different science teams to prepare their
individual analysis tools on data that would be as close to the final data
products as feasible while being able to generate these mock data in a
reasonable amount of time on their local desktop machines.
An alternative tool, called the Instrument Numerical Model
(INM)\cite{JBF+08,Jarno+10}, was able to simulate raw MUSE calibration and
science exposures, but required computations of several days on a powerful
workstation to generate the data.  These in turn then had to be reduced using
the data reduction pipeline to create datacubes.
Compared to this, the QSIM datacube simulator was a very lightweight tool to
generate datacubes quickly.

In 2013 QSIM was rewritten by M.\ Wendt from scratch in Python 2.X on the basis
of the MUSE Python Data Analysis Framework (MPDAF) developed at the same time
in Lyon \cite{2016ascl.soft11003B}.
In the end only few science simulations were realized before the First Light of
MUSE itself.
Namely, mockup cubes of parts of the UDF fields as well as a comprehensive
simulation of a globular cluster field containing about 38,000 stars that also
served as testing ground for resolving stellar populations with crowded field
3D spectroscopy\cite{KWR13}.
Fortunately, the MUSE instrument was very successful right from the start and
even the commissioning data led to science publications. In particular of the
globular cluster NGC\,6397, featuring the aforementioned crowded field
spectroscopy\cite{2016A&A...588A.148H} and a transverse science case directly
motivated by the earlier simulations\cite{2017A&A...607A.133W}. Once real
instrument data was available, there was little use for a simulation tool in
its state at the time.

\subsection{Development of simulation software for BlueMUSE}
During the preparation for the upcoming MUSE data before commissioning, it
became obvious that the future end users of the final data products needed to
get ready to deal with that kind of data in terms of sheer volume and also the
limited ability of common tools to handle 3-dimensional data sets.
In contrast, future users of BlueMUSE will be able to draw from a large pool of
released tools and experience as well as laptop sized hardware that is easily
up to the task of managing the expected data end products.
A lesson learned from the development of dedicated simulation tools is that
they need to be available way ahead of the planning of science observation
since this is where simulations can play out their full potential.
Accessing feasiblity and requirements for certain scientific questions in
relation to a group of objects.
While software like the provided exposure time calculator (ETC) will provide
estimates of the expected signal-to-noise for a given observation, its impact
on the success of the scientific exploitation of the data can only be tested
and confirmed deploying the full analysis chain on the data.
A number -- but not all -- of the questions that can and should be tackled ahead
of the observation proposals are:
\begin{itemize}
\item What is the minimum S/N at which we can expect to detect the
      characteristic features of interest?
\item Is the given resolution as well as the binsize appropriate for the set
      goals?
\item How much is the data being affected by the (reduced) transmission of the
      instrument (at the lower wavelength end)?
\item Are sky emission and telluric absorption lines a problem for the specific
      redshift range?
\end{itemize}

To successfully tackle those questions, the simulations need to create as
realistic as possible data that mimics the expected data product from the
offical data reduction pipeline closely.
This includes the geometrical properties of the data cube and the plain
instrumental parameters such as the spatial and spectral sampling, the spectral
resolution as well as the overall cube dimensions as mentioned in section
\ref{sec:intro}.
Naturally it needs an interface to implement various science objects in the
simulations, ranging from individual point sources such as quasars to clusters
of stars, groups of galaxies or even diffuse emission.

But it also has to implement the simulated observing conditions, most notably
the seeing conditions but also the full composition of the telluric background
and the resulting complex noise properties of the data.
The infamous 'redshift desert' for example -- the inability or lowered
sensitivity to appropiately determine galaxy redshifts in a certain redshift
range (usually $1.4 < z < 1.9$) -- can in part be attributed to the lack of
suitable emission lines in the optical for certain redshift ranges but a major
aspect is the contamination with telluric features for higher wavelength
ranges. Even single features can have influence on detection software, which is
only one aspect why an accurate sky modelling is fundamental.

To address the mentioned points, a completely new software is currently being
developed at Potsdam University to simulate BlueMUSE data: 'BlueSi'.  Where the
former QSIM relied on frequent changes of an early MPDAF development version
and on Python 2.x the focus is now on user-friendlyness and long-term
maintainability to have a ready-to-use product in Python 3.x already in the
science planning phase of the instrument.
In the current early state, the BlueMUSE simulation software 'BlueSi'
implements a Python interface to The Cerro Paranal Advanced Sky
Model\footnote{This model is available from
  \url{https://www.eso.org/sci/software/pipelines/skytools/skymodel}.}
which allows the user to either manually vary up to 36 parameters related to
physical conditions or simply mimic the conditions at a specific day and time
and optionally merely provide deviations to those. This will allow the user to
make very detailed choices while also being able to rely on realistic default
settings.

The latter point turned out to be crucial in the usability and overall value of
such simulations software and it also affects the inclusion of science objects
into the simulated data cubes.
While the simulation software for MUSE allowed for very complex and specific
objects such as galaxies with separated information on different stellar
populations with their own distributions and kinematics along with parametrized
rotational curves and dispersion maps of any number of emission lines, it was
basically impossible for a non-expert in all of those fields to generate
realistic representations of typical galaxies.
The mere preparation of the source file in the proper format constituted a
significant obstacle.

To provide a much smoother learning curve and ease the process of simulating
data products for a range of science cases, we work in close contact with the
potential future users to help create an intuitive user interface and data
format per major science case. We hope that by establishing a range of actual
science objects as provided by the related peer-group the threshold to derive
new science objects as targets for the BlueSi software will be significantly
lowered.

As part of the current efforts, we implement a simulation of a globular cluster
based on the data and analysis of MUSE observations.
The detailed study of NGC\,3201\cite{2018MNRAS.475L..15G} allows us to to make
use of a large data base of synthetic stellar spectra
(PHOENIX\cite{2013A&A...553A...6H}) based on the derived stellar parameters.
These synthetic spectra cover the full wavelength range of BlueMUSE and are
computed at very high resolution ($R\sim500,000$). Naturally, the simulation
also needs to incorporate the features related to the cluster itself, such as
the individual radial velocities of the stars and their intrinsic magnitudes.

The different steps involved in defining the data format and interface to the
simulation software will be documented and serve as a blueprint for further
integration.

\section{CONCLUSIONS AND OUTLOOK}\label{sec:conc}
Despite some issues, the MUSE integral field spectrograph and its pipeline are
a major success. The pipeline was already in a good shape to support
commissioning activities in 2014 and first science papers shortly thereafter.
Most of the shortcomings of the pipeline are understood and will be rectified
in the design of the BlueMUSE pipeline, with MUSE data and pipeline as its
testbed.

Specific improvements planned for the BlueMUSE data reduction are a simpler
structure with better parallelization, improved wavelength calibration through
Fabry-Perot or frequency comb techniques, a more detailed knowledge of the
line-spread function, and the propagation of covariance to the final datacube.

To support science studies, a new simulation suite 'BlueSi' is being developed,
with globular clusters as the first case to test generation of a realistic
datacube.

The BlueMUSE integral field spectrograph is currently in its internal Phase A
study phase and will enter the formal process with ESO in 2024. It is planned
to be installed at the VLT in 2030 by which time its pipeline should be largely
finished and ready for testing on sky.

\acknowledgments
PMW, SM, MMR, and AK gratefully acknowledge support by the BMBF from the ErUM
program (project VLT-BlueMUSE, grants 05A20BAB and 05A20MGA).

\bibliography{spie}

\begin{thebibliography}{10}

\bibitem{Bacon+10}
{Bacon}, R., {Accardo}, M., {Adjali}, L., {Anwand}, H., {Bauer}, S., {et~al.},
  ``{The MUSE second-generation VLT instrument},'' in [{\em {Ground-based and
  Airborne Instrumentation for Astronomy III}}{\nolinebreak\hspace{0.1em}]},
  {\em Proc.~{SPIE}} {\bf 7735} (July 2010).

\bibitem{2014Msngr.157...13B}
Bacon, R., Vernet, J., Borisova, E., Bouch\'e, N., Brinchmann, J., {et~al.},
  ``{MUSE Commissioning},'' {\em The Messenger}~{\bf 157},  13--16 (Sept.
  2014).

\bibitem{BlueMUSE_1906.01657}
{Richard}, J., {Bacon}, R., {Blaizot}, J., {Boissier}, S., {Boselli}, A.,
  {et~al.}, ``{BlueMUSE: Project Overview and Science Cases},'' {\em arXiv}
  (Jun 2019).
\newblock {BlueMUSE white paper, arXiv:1906.01657}.

\bibitem{musepipeline}
{Weilbacher}, P.~M., {Palsa}, R., {Streicher}, O., {Bacon}, R., {Urrutia}, T.,
  {et~al.}, ``{The Data Processing Pipeline for the MUSE Instrument},'' {\em
  A\&A}~{\bf 641},  A28 (Sept. 2020).

\bibitem{2019Icar..331...69I}
{Irwin}, P. G.~J., {Toledo}, D., {Braude}, A.~S., {Bacon}, R., {Weilbacher},
  P.~M., {et~al.}, ``{Latitudinal variation in the abundance of methane
  (CH$_{4}$) above the clouds in Neptune's atmosphere from VLT/MUSE Narrow
  Field Mode Observations},'' {\em Icarus}~{\bf 331},  69--82 (Oct 2019).

\bibitem{2019A&A...628A.128O}
{Opitom}, C., {Yang}, B., {Selman}, F., and {Reyes}, C., ``{First observations
  of an outbursting comet with the MUSE integral-field spectrograph},'' {\em
  A\&A}~{\bf 628},  A128 (Aug. 2019).

\bibitem{McLeod15b}
{McLeod}, A.~F., {Weilbacher}, P.~M., {Ginsburg}, A., {Dale}, J.~E., {Ramsay},
  S., and {Testi}, L., ``{A nebular analysis of the central Orion nebula with
  MUSE},'' {\em MNRAS}~{\bf 455},  4057--4086 (Feb. 2016).

\bibitem{2020A&A...634A..47M}
{Monreal-Ibero}, A. and {Walsh}, J.~R., ``{The MUSE view of the planetary
  nebula NGC 3132},'' {\em A\&A}~{\bf 634},  A47 (Feb. 2020).

\bibitem{2018MNRAS.473.5591K}
{Kamann}, S., {Husser}, T.~O., {Dreizler}, S., {Emsellem}, E., {Weilbacher},
  P.~M., {et~al.}, ``{A stellar census in globular clusters with MUSE: The
  contribution of rotation to cluster dynamics studied with 200 000 stars},''
  {\em MNRAS}~{\bf 473},  5591--5616 (Feb 2018).

\bibitem{2019A&A...632A...3G}
{Giesers}, B., {Kamann}, S., {Dreizler}, S., {Husser}, T.-O., {Askar}, A.,
  {et~al.}, ``{A stellar census in globular clusters with MUSE: Binaries in NGC
  3201},'' {\em A\&A}~{\bf 632},  A3 (Dec. 2019).

\bibitem{2018A&A...618A...3R}
{Roth}, M.~M., {Sandin}, C., {Kamann}, S., {Husser}, T.-O., {Weilbacher},
  P.~M., {et~al.}, ``{MUSE crowded field 3D spectroscopy in NGC 300. I. First
  results from central fields},'' {\em A\&A}~{\bf 618},  A3 (Oct. 2018).

\bibitem{2020A&A...633A..79K}
{Kollatschny}, W., {Weilbacher}, P.~M., {Ochmann}, M.~W., {Chelouche}, D.,
  {Monreal-Ibero}, A., {et~al.}, ``{NGC 6240: A triple nucleus system in the
  advanced or final state of merging},'' {\em A\&A}~{\bf 633},  A79 (Jan.
  2020).

\bibitem{2022A&A...659A.191E}
{Emsellem}, E., {Schinnerer}, E., {Santoro}, F., {Belfiore}, F., {Pessa}, I.,
  {et~al.}, ``{The PHANGS-MUSE survey. Probing the chemo-dynamical evolution of
  disc galaxies},'' {\em A\&A}~{\bf 659},  A191 (Mar. 2022).

\bibitem{2018Natur.562..229W}
{Wisotzki}, L., {Bacon}, R., {Brinchmann}, J., {Cantalupo}, S., {Richter}, P.,
  {et~al.}, ``{Nearly all the sky is covered by Lyman-{\ensuremath{\alpha}}
  emission around high-redshift galaxies},'' {\em Nature}~{\bf 562},  229--232
  (Oct 2018).

\bibitem{2021A&A...647A.107B}
{Bacon}, R., {Mary}, D., {Garel}, T., {Blaizot}, J., {Maseda}, M., {et~al.},
  ``{The MUSE Extremely Deep Field: The cosmic web in emission at high
  redshift},'' {\em A\&A}~{\bf 647},  A107 (Mar. 2021).

\bibitem{2014MNRAS.445L..79E}
{Emsellem}, E., {Krajnovic}, D., and {Sarzi}, M., ``{A kinematically distinct
  core and minor-axis rotation: the MUSE perspective on M87.},'' {\em
  MNRAS}~{\bf 445},  L79--L83 (Nov. 2014).

\bibitem{2014MNRAS.445.4335F}
{Fumagalli}, M., {Fossati}, M., {Hau}, G. K.~T., {Gavazzi}, G., {Bower}, R.,
  {et~al.}, ``{MUSE sneaks a peek at extreme ram-pressure stripping events - I.
  A kinematic study of the archetypal galaxy ESO137-001},'' {\em MNRAS}~{\bf
  445},  4335--4344 (Dec. 2014).

\bibitem{2016MNRAS.458.3210S}
{Soto}, K.~T., {Lilly}, S.~J., {Bacon}, R., {Richard}, J., and {Conseil}, S.,
  ``{ZAP - enhanced PCA sky subtraction for integral field spectroscopy},''
  {\em MNRAS}~{\bf 458},  3210--3220 (May 2016).

\bibitem{2019MNRAS.483.5188C}
{Cantalupo}, S., {Pezzulli}, G., {Lilly}, S.~J., {Marino}, R.~A., {Gallego},
  S.~G., {et~al.}, ``{The large- and small-scale properties of the
  intergalactic gas in the Slug Ly {\ensuremath{\alpha}} nebula revealed by
  MUSE He II emission observations},'' {\em MNRAS}~{\bf 483},  5188--5204 (Mar.
  2019).

\bibitem{2022A&A...659A.124H}
{Husemann}, B., {Singha}, M., {Scharw{\"a}chter}, J., {McElroy}, R., {Neumann},
  J., {et~al.}, ``{The Close AGN Reference Survey (CARS). IFU survey data and
  the BH mass dependence of long-term AGN variability},'' {\em A\&A}~{\bf 659},
   A124 (Mar. 2022).

\bibitem{2015A&A...575A..75B}
{Bacon}, R., {Brinchmann}, J., {Richard}, J., {Contini}, T., {Drake}, A.,
  {et~al.}, ``{The MUSE 3D view of the Hubble Deep Field South},'' {\em
  A\&A}~{\bf 575},  A75 (Mar. 2015).

\bibitem{2019A&A...624A.141U}
{Urrutia}, T., {Wisotzki}, L., {Kerutt}, J., {Schmidt}, K.~B., {Herenz}, E.~C.,
  {et~al.}, ``{The MUSE-Wide Survey: survey description and first data
  release},'' {\em A\&A}~{\bf 624},  A141 (Apr 2019).

\bibitem{MUSE_User_Manual_v10.4}
Richard, J., Bacon, R., Vernet, J., Wylezalek, D., Valenti, E., {et~al.}, {\em
  MUSE User Manual}.
\newblock ESO (Aug. 2019).
\newblock {ESO-261650}, v10.4.

\bibitem{2017A&A...608A...1B}
{Bacon}, R., {Conseil}, S., {Mary}, D., {Brinchmann}, J., {Shepherd}, M.,
  {et~al.}, ``{The MUSE Hubble Ultra Deep Field Survey. I. Survey description,
  data reduction, and source detection},'' {\em A\&A}~{\bf 608},  A1 (Nov
  2017).

\bibitem{2015ascl.soft04003E}
{ESO CPL Development Team}, ``{EsoRex: ESO Recipe Execution Tool}.''
  Astrophysics Source Code Library (Apr. 2015).
\newblock ascl:1504.003.

\bibitem{2022SPIE_EDPS}
{Freudling}, W., ``{The ESO Data Processing System (EDPS): A unified system for
  science data processing},'' in [{\em
  Proc.~SPIE}{\nolinebreak\hspace{0.1em}]},   {\bf 12186},  1218612 (2022).

\bibitem{molecfit}
{Smette}, A., {Sana}, H., {Noll}, S., {Horst}, H., {Kausch}, W., {et~al.},
  ``{Molecfit: A general tool for telluric absorption correction. I. Method and
  application to ESO instruments},'' {\em A\&A}~{\bf 576},  A77 (Apr. 2015).

\bibitem{2015ascl.soft01013S}
{Smette}, A., {Kausch}, W., {Sana}, H., {Noll}, S., {Horst}, H., {et~al.},
  ``{Molecfit: Telluric absorption correction tool}.'' Astrophysics Source Code
  Library (Jan. 2015).

\bibitem{1996ASPC..101..207W}
{Wallace}, P.~T., ``{The IAU SOFA Initiative},'' in [{\em Astronomical Data
  Analysis Software and Systems V}{\nolinebreak\hspace{0.1em}]},  {Jacoby},
  G.~H. and {Barnes}, J., eds., {\em Astronomical Society of the Pacific
  Conference Series} {\bf 101},  207 (Jan. 1996).

\bibitem{2014ascl.soft03026I}
{IAU SOFA Center}, ``{SOFA: Standards of Fundamental Astronomy}.'' Astrophysics
  Source Code Library, record ascl:1403.026 (Mar. 2014).

\bibitem{2022SPIE_BlueMUSE_CU}
{Roth}, M.~M., {Kelz}, A., {Madhav}, K., {Weilbacher}, P.~M., {Richard}, J.,
  {et~al.}, ``{The BlueMUSE Calibration Unit: Phase-A studies},'' in [{\em
  Proc.~SPIE}{\nolinebreak\hspace{0.1em}]},   {\bf 12184},  12184--217 (2022).

\bibitem{2003MNRAS.342..345C}
{Cappellari}, M. and {Copin}, Y., ``{Adaptive spatial binning of integral-field
  spectroscopic data using Voronoi tessellations},'' {\em MNRAS}~{\bf 342},
  345--354 (June 2003).

\bibitem{2017MNRAS.466..798C}
{Cappellari}, M., ``{Improving the full spectrum fitting method: accurate
  convolution with Gauss-Hermite functions},'' {\em MNRAS}~{\bf 466},  798--811
  (Apr. 2017).

\bibitem{2016AJ....152...83L}
{Law}, D.~R., {Cherinka}, B., {Yan}, R., {Andrews}, B.~H., {Bershady}, M.~A.,
  {et~al.}, ``{The Data Reduction Pipeline for the SDSS-IV MaNGA IFU Galaxy
  Survey},'' {\em AJ}~{\bf 152},  83 (Oct. 2016).

\bibitem{JBF+08}
{Jarno}, A., {Bacon}, R., {Ferruit}, P., and {P{\'e}contal-Rousset}, A.,
  ``{Numerical Simulation of the VLT/MUSE Instrument},'' in [{\em Astronomical
  {D}ata {A}nalysis {S}oftware and {S}ystems
  {XVII}}{\nolinebreak\hspace{0.1em}]},  {Argyle}, R.~W., {Bunclark}, P.~S.,
  and {Lewis}, J.~R., eds., {\em ASP Conf. Ser.} {\bf 394},  701 (Aug. 2008).

\bibitem{Jarno+10}
{Jarno}, A., {Bacon}, R., {Ferruit}, P., {P{\'e}contal-Rousset}, A.,
  {Pandey-Pommier}, M., {et~al.}, ``{Introducing atmospheric effects in the
  numerical simulation of the VLT/MUSE instrument},'' in [{\em {Modeling,
  Systems Engineering, and Project Management for Astronomy
  IV}}{\nolinebreak\hspace{0.1em}]},  {\em Proc.~{SPIE}} {\bf 7738} (July
  2010).

\bibitem{2016ascl.soft11003B}
{Bacon}, R., {Piqueras}, L., {Conseil}, S., {Richard}, J., and {Shepherd}, M.,
  ``{MPDAF: MUSE Python Data Analysis Framework}.'' Astrophysics Source Code
  Library (Nov 2016).
\newblock ascl:1611.003.

\bibitem{KWR13}
{Kamann}, S., {Wisotzki}, L., and {Roth}, M.~M., ``{Resolving stellar
  populations with crowded field 3D spectroscopy},'' {\em A\&A}~{\bf 549},  A71
  (Jan. 2013).

\bibitem{2016A&A...588A.148H}
{Husser}, T.-O., {Kamann}, S., {Dreizler}, S., {Wendt}, M., {Wulff}, N.,
  {et~al.}, ``{MUSE crowded field 3D spectroscopy of over 12 000 stars in the
  globular cluster NGC 6397. I. The first comprehensive HRD of a globular
  cluster},'' {\em A\&A}~{\bf 588},  A148 (Apr. 2016).

\bibitem{2017A&A...607A.133W}
{Wendt}, M., {Husser}, T.-O., {Kamann}, S., {Monreal-Ibero}, A., {Richter}, P.,
  {et~al.}, ``{Mapping diffuse interstellar bands in the local ISM on small
  scales via MUSE 3D spectroscopy. A pilot study based on globular cluster NGC
  6397},'' {\em A\&A}~{\bf 607},  A133 (Nov. 2017).

\bibitem{2018MNRAS.475L..15G}
{Giesers}, B., {Dreizler}, S., {Husser}, T.-O., {Kamann}, S., {Anglada
  Escud{\'e}}, G., {et~al.}, ``{A detached stellar-mass black hole candidate in
  the globular cluster NGC 3201},'' {\em MNRAS}~{\bf 475},  L15--L19 (Mar.
  2018).

\bibitem{2013A&A...553A...6H}
{Husser}, T.~O., {Wende-von Berg}, S., {Dreizler}, S., {Homeier}, D.,
  {Reiners}, A., {et~al.}, ``{A new extensive library of PHOENIX stellar
  atmospheres and synthetic spectra},'' {\em A\&A}~{\bf 553},  A6 (May 2013).

\end{thebibliography}
\bibliographystyle{spiebib}

\end{document}